\documentclass[12pt]{article}
\usepackage{amsmath, amssymb}
\usepackage[dvips]{epsfig}

\def\##1{\underline #1}
\def\=#1{\underline{\underline #1}}
\def\4#1{\underline{\underline{\underline{\underline #1}}}}

\def\.{\mbox{ \tiny{$^\bullet$} }}

\def\les{\left[}
\def\ris{\right]}
\def\lec{\left\{}
\def\ric{\right\}}

\def\c#1{\cite{#1}}

\def\r#1{(\ref{#1})}

\def\curl{\nabla\times}

\def\eps{\epsilon}
\def\epso{\eps_0}
\def\muo{\mu_0}

\date{}

\begin{document}

\baselineskip .582cm   

\begin{center}
\vskip 0.2cm
{\large\bf ON PERFECT LENSES AND NIHILITY}\\[20pt]
{\bf Akhlesh Lakhtakia}
\\[20pt]
{\em CATMAS -- Computational \& Theoretical Materials Sciences Group\\
Department of Engineering Science and Mechanics\\
Pennsylvania State University, University Park, PA 16802--6812, USA}\\[25pt]
\end{center}

\begin{abstract}
The canonical problem of a perfect lens with linear bianisotropic
mater\-ials is formulated. Its solution is shown
to be directly connected with
the concept of nihility, the electromagnetic nilpotent. Perfect lenses as well as
nihility remain unrealizable.
\\

KEYWORDS: Anti--vacuum, Lens, Negative permeability, 
Negative permittivity,   Nihility, Perfect lens
\end{abstract}

\section{Introduction}
This communication has been inspired by a report on the
theory of a perfect lens by Pendry \c{P}. This lens
is supposedly constructed of a material whose permittivity
and permeability, respectively, are exactly negative of the
permittivity and the permeability of free space. In  
 a predecessor paper \c{Lakh1},
that material was postulated as the {\em anti--vacuum\/}.

\section{Canonical Formulation for a Lens}
As geometrical optics is applicable for lenses, the canonical
formulation  for a lens merely involves a linear homogeneous material
confined to the region between two parallel planes.
Let us, however, generalize the situation in order to understand
the issue at greater depth by involving two linear homogeneous
materials and four interfaces, as shown in Figure 1. The regions
$0 \leq z \leq d_1$ and $d_1+d_2 \leq z \leq d_1+d_2+d_3$
are occupied by a material labelled $a$, and the region
$d_1 \leq z \leq d_1+d_2$ by a material labelled $b$.
Both materials are linear, homogeneous, bianisotropic
and necessarily dispersive, their frequency--domain constitutive relations being as follows:
\begin{equation}
\left. \begin{array}{l}
{\bf D}(x,y,z,\omega) = \epso\, \les \=\eps^{a,b}(\omega)\.{\bf E}\,(x,y,z,\omega) +
\=\alpha^{a,b}(\omega)\.{\bf H}\,(x,y,z,\omega)\ris\\[10pt]
{\bf B}(x,y,z,\omega) = \muo\, \les \=\beta^{a,b}(\omega)\.{\bf E}\,(x,y,z,\omega) +
\=\mu^{a,b}(\omega)\.{\bf H}\,(x,y,z,\omega)\ris
\end{array}\ric\,.
\end{equation}
In these relations, the dielectric properties are delineated by 
$\=\eps^{a,b}(\omega)$,
the magnetic properties by $\=\mu^{a,b}(\omega)$, and the magnetoelectric properties
by $\=\alpha^{a,b}(\omega)$ as well as $\=\beta^{a,b}(\omega)$, all of these
dyadics being functions of the angular frequency $\omega$.
The permittivity and permeability of free space (i.e., vacuum) are denoted
by $\epso$ and $\muo$, respectively.

\begin{center}
\begin{figure}[!ht]
\centering \psfull \epsfig{file=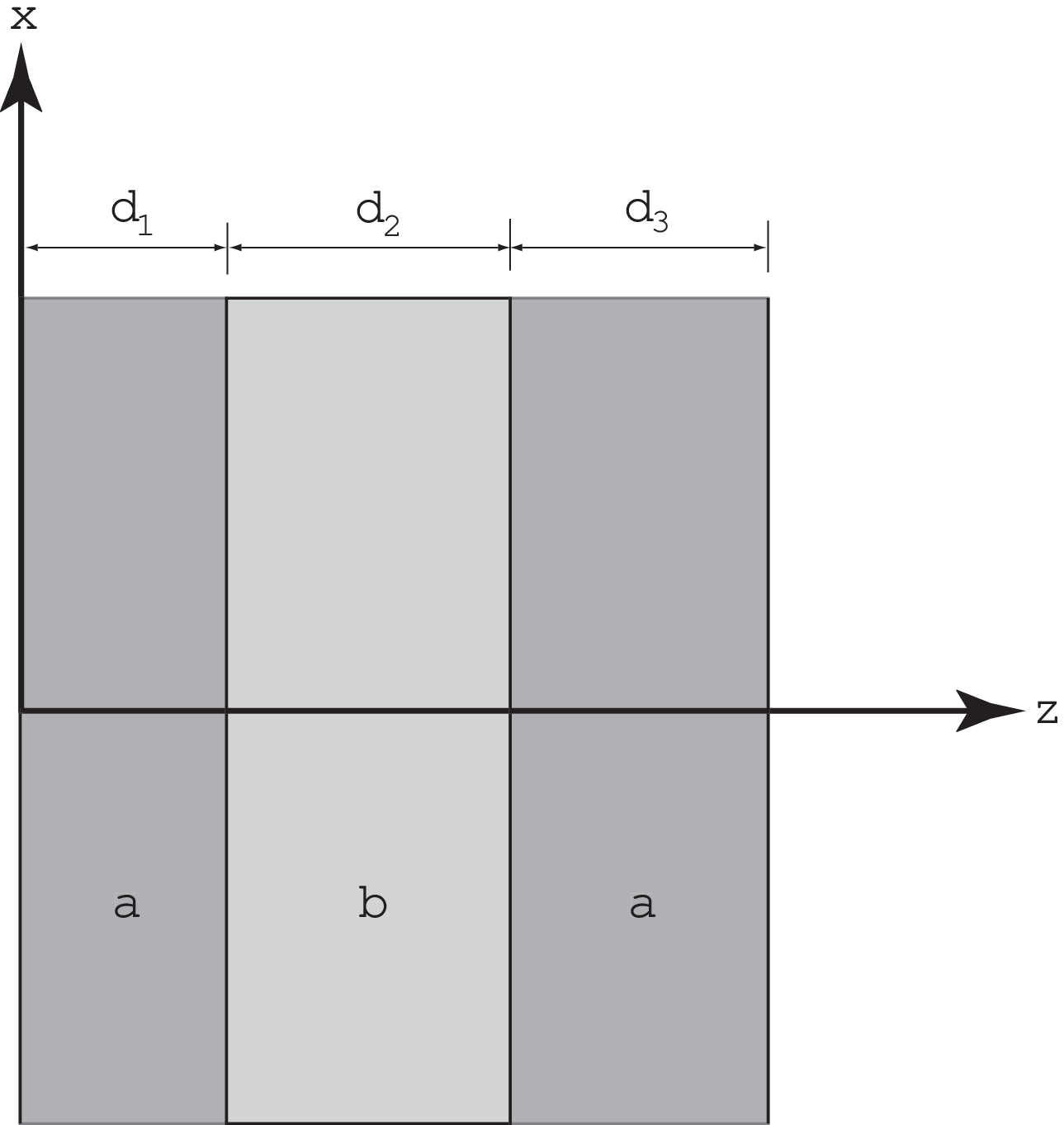,width=3.1in}
\end{figure}
\bigskip 
{\bf Figure 1:}
Schematic for the canonical lens formulation.
\end{center}
\bigskip

Without loss of essential generality, we can take the spatial Fourier transformations
of all electromagnetic phasors with respect to $x$ and $y$; thus,
\begin{equation}
{\bf E}(x,y,z,\omega) = {\bf e}(z,\kappa,\psi,\omega) \exp\les i\kappa
(x\cos\psi + y\sin\psi)\ris\,,
\end{equation}
etc., suffice for the present purposes. Wave propagation in the two materials
can then be cast in terms of 4$\times$4 matrix ordinary differential equations
as follows \c{Lakh2}:
\begin{equation}
\frac{d}{dz}\,[\#f(z,\kappa,\psi,\omega)]= i [\=P_{\,a,b}(\kappa,\psi,\omega)]\.
[\#f(z,\kappa,\psi,\omega)]\,.
\end{equation}
Here, $[\#f] \equiv [e_x;\,e_y;\,h_x;\,h_y]^T$ is a column 4--vector
with the superscript $^T$ denoting the transpose, while
$[\=P_{\,a,b}]$ are 4$\times$4 matrixes. Accordingly,
we obtain the basic relation
\begin{equation}
[\#f(d_1+d_2+d_3,\kappa,\psi,\omega)]= [\=M(d_1+d_2+d_3,\kappa,\psi,\omega)]\.
[\#f(0,\kappa,\psi,\omega)]\,,
\end{equation}
where
\begin{eqnarray}
\nonumber
[\=M(d_1+d_2+d_3,\kappa,\psi,\omega)]&=& \exp\lec id_3[\=P_{\,a}(\kappa,\psi\,\omega)]\ric\.
\exp\lec id_2[\=P_{\,b}(\kappa,\psi\,\omega)]\ric\\
&&\quad \.
\exp\lec id_1[\=P_{\,a}(\kappa,\psi\,\omega)]\ric\,.
\end{eqnarray}

Within the confines of continuum electromagnetics,
the canonical lens problem  involves finding the material $b$ and the
thickness $d_2$ for a specified material $a$ and thicknesses $d_1$ and $d_3$,
such that
\begin{equation}
\label{lens}
[\=M(d_1+d_2+d_3,\kappa,\psi,\omega)]= 
\les\begin{array}{cccc}
1 & 0 & 0 & 0 \\ 0 & 1 & 0 & 0 \\ 0 & 0 & 1 & 0 \\ 0 & 0 & 0 & 1
\end{array}\ris
\end{equation}
for {\em all\/} $\kappa$, $\psi$ and $\omega$.

\section{Analysis and Discussion}
Obviously, that would be a fruitless endeavor in practice. Hence, lens designers 
{\em effectively\/} settle
for some acceptable ranges of $\kappa$, $\psi$ and $\omega$
in which \r{lens} holds, when material $a$ is air. Deviations from
an ideal match introduce aberrations \c{Sin}.

Mathematically, and at first glance, an excellent
candidate for ideal
match is the following:
\begin{equation}
\label{ideal}
\left.\begin{array}{ll}
\=\eps^{b}(\omega) = -\=\eps^{a}(\omega)\,, &\quad
\=\alpha^{b}(\omega) = -\=\alpha^{a}(\omega)\, \\
\=\beta^{b}(\omega) = -\=\beta^{a}(\omega)\,, &\quad
\=\mu^{b}(\omega) = -\=\mu^{a}(\omega)\, \\
d_2=d_1+d_3
\end{array}\ric\,.
\end{equation}
Close inspection of \r{lens}, however, shows that \r{ideal}
is suitable for all $\kappa$,
only if material $a$ has orthorhombic symmetry, i.e., 
\begin{equation}
\label{ideal1}
\=\xi^{a} = \les \begin{array}{ccc}
\xi_1^a & 0 & 0 \\ 0 & \xi_2^a & 0 \\ 0 & 0 & \xi_3^a
\end{array}\ris\,,\qquad \xi = \eps\,,\,\alpha\,,\beta\,,\mu\,.
\end{equation}

A sandwich of equal thicknesses of materials $a$ and $b$
then constitutes a planar realization of a medium named {\em nihility}
earlier in this journal \c{Lakh1}. Nihility is the postulated electromagnetic nilpotent, with the following
constitutive relations:
\begin{equation}
\left. \begin{array}{l}
{\bf D}(x,y,z,\omega) = {\bf 0} \\[10pt]
{\bf B}(x,y,z,\omega) = {\bf 0}
\end{array}\ric\,.
\end{equation}
Wave propagation cannot occur in nihility, because $\curl {\bf E}(x,y,z,\omega) ={\bf 0}$
and $\curl {\bf H}(x,y,z,\omega)  = {\bf 0}$ in the absence of sources therein.
Whereas the phase velocity and the
wavevector
of a plane wave in vacuum/anti--vacuum are co--parallel/anti--parallel,
the directionality of the phase velocity relative to the
wave\-vector in nihility is a non--issue.

Physically, \r{ideal} and \r{ideal1} are still deficient because
the principle of energy conservation has
not been considered. With the normal constraint that material $a$ be
passive, it follows from \r{ideal} that material $b$ has to be active.
Although their electromagnetic response  can be simulated {\em via\/} composites
containing
active circuit elements \c{AZ},
I do not think that active materials will provide a realistic option 
in the near future. Therefore, the passivity constraint on material
$b$ would lead to aberrations due to absorption. To those must be
added chromatic aberrations, which would arise from the non--fulfilment
of \r{ideal} outside some limited range of $\omega$.

Suppose next that material $a$ is isotropic and non--magnetoelectric, i.e.,
$\=\eps^a=\eps^a\,\=I$, $\=\mu^a=\mu^a\,\=I$, and $\=\alpha^a=\=\beta^a = \=0$,
where $\=I$ is the identity dyadic and $\=0$ is the null dyadic. Then, a reasonable match
in some frequency range could conceivably be
provided by materials that supposedly possess a negative index of refraction \c{SSS},
so long as absorption is acceptably low.

Pendry \c{P} actually took material $a$ to be air, which
is optically indistinguishable from vacuum ($\=\eps^a= \=I$, $\=\mu^a= \=I$, and $\=\alpha^a=\=\beta^a = \=0$)
for most purposes. Thus, the requirements on material
$b$ became very simple, {\em viz.\/}, $\=\eps^b=- \=I$, $\=\mu^b= -\=I$, and $\=\alpha^b=\=\beta^b = \=0$. (In other words,
material $b$ has to be anti--vacuum.) But
Ziolkowski \c{Z} has recently concluded from
two--dimensional computer simulations
that even these simple
requirements (in some narrow frequency range) cannot be met by realistic meta--materials. 
Recent correspondence between Pendry and others adds to the debate \c{P1}--\c{P4}.

A perfect lens
remains unrealizable in my opinion, as does nihility.

\small{

}

\end{document}